\documentclass[aip,apl,reprint,amsmath,amssymb,twocolumn]{revtex4-1}

\usepackage{graphicx}
\usepackage{hyphenat}
\usepackage{siunitx}

\usepackage[caption=false]{subfig} 

\usepackage{color}

\usepackage{natbib}
\usepackage{siunitx}
\usepackage{amsmath}

\usepackage{color}

\begin{document}

\title{Mechanically tunable integrated beamsplitters on a flexible polymer platform}

\author{James A. Grieve}
\email{james.grieve@nus.edu.sg}
\author{Kian Fong Ng}
\affiliation{Centre for Quantum Technologies, National University of Singapore, Blk S15, 3 Science Drive 2, 117543 Singapore}

\author{Manuel J. L. F. Rodrigues}
\affiliation{Centre for Advanced 2D Materials and Graphene Research Centre, National University of Singapore, 6 Science Drive 2, Singapore 117546, Singapore}

\author{Jos\'{e} Viana-Gomes}
\affiliation{Centre for Advanced 2D Materials and Graphene Research Centre, National University of Singapore, 6 Science Drive 2, Singapore 117546, Singapore}
\affiliation{Department of Physics, National University of Singapore, Blk S12, 2 Science Drive 3, 117551 Singapore}

\author{Alexander Ling}
\affiliation{Centre for Quantum Technologies, National University of Singapore, Blk S15, 3 Science Drive 2, 117543 Singapore}
\affiliation{Department of Physics, National University of Singapore, Blk S12, 2 Science Drive 3, 117551 Singapore}

\date{\today}

\begin{abstract}
We report the development of a monolithic, mechanically tunable waveguide platform based on the flexible polymer polydimethyl siloxane (PDMS). Such devices preserve single mode guiding across a wide range of linear geometric distortions. This enables the realization of directional couplers with tunable splitting ratios via elastic deformation of the host chip. We fabricated several devices of this type, and verified their operation over a range of wavelengths, with access to the full range of input/output ratios. The low cost and relative ease of fabrication of these devices via a modified imprint lithographic technique make them an attractive platform for investigation of large scale optical random walks and related optical phenomena.
\end{abstract}

\maketitle



Integrated optical systems based on waveguide platforms have become widespread in both industrial and research settings. While the ability to mass produce an optical circuit with a very small footprint has doubtless been a major contributor to their success, the greatest advantage conferred by an integrated rather than bulk optics approach is that of stability. In an industrial setting this has enabled the deployment of complex optical systems in areas where servicing and alignment would have been impossible, while in the research community it has enabled the realization of large scale optical circuits which would have been prohibitively difficult to align and maintain using macroscopic components.

In the field of quantum optics, integrated photonics technologies have yielded quantum random walks with over 100 optical modes \cite{Perets2008,Peruzzo2010}, as well as functional circuits containing a very large number of linear optical elements \cite{Crespi2011,Li2011,Politi2009} and photonic simulators with a large number of modes \cite{Zeuner2012,Plotnik2014}. In general, however, these experiments were unable to leverage the available commercial platforms due to physical limitations on permitted wavelength and polarization modes. For example, silicon-based platforms preclude the use of wavelengths below 1 micron where good low-cost single photon detectors are available. Consequently many prominent results have been realized using bespoke platforms (such as laser-inscribed waveguides in glass \cite{Plotnik2014}) which lack the extensive development present in their commercial counterparts. This comparative immaturity is felt most keenly when elements of an optical circuit must be reconfigurable.

Tunability of integrated photonics platforms is typically achieved either by accessing the host material's electro-optic response (for example lithium niobate\cite{Alferness1978}), or more commonly, by making use of thermal effects. In particular, the leading commercial platforms can achieve high performance thermal tuning by leveraging on significant industry-led engineering efforts~\cite{Notaros2017}. 


\begin{figure}[b]
	\includegraphics[width=\linewidth]{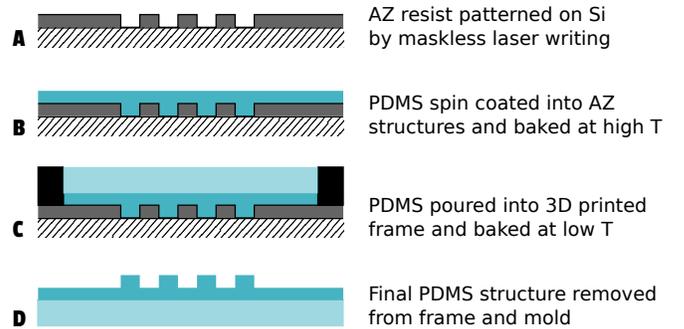}
    \caption{\label{fig:fabrication}Polydimethyl siloxane (PDMS) chips are fabricated by a simple molding technique. An AZ photoresist (AZ1512HS, AZ Electronic Materials) is patterned onto a Silicon wafer using a maskless laser write system. Liquid PDMS prepared according to the manufacturer's instructions is spin coated into the AZ mold at \SI{6000}{RPM} for 10 minutes, overfilling the features with a \SI{4}{\micro\meter} layer. This structure is cured at \SI{150}{\celsius} for 60 minutes to produce high density PDMS (HD-PDMS). After cooling, a 3D printed frame is applied and a thick (\SIrange{1}{2}{mm}) layer of PDMS is poured onto the HD-PDMS. After curing at \SI{70}{\celsius} for 60 minutes, the PDMS chip is removed from the frame and AZ mold, exposing the final pattern.}
\end{figure}

\begin{figure*}[t]
	\includegraphics[width=0.9\linewidth]{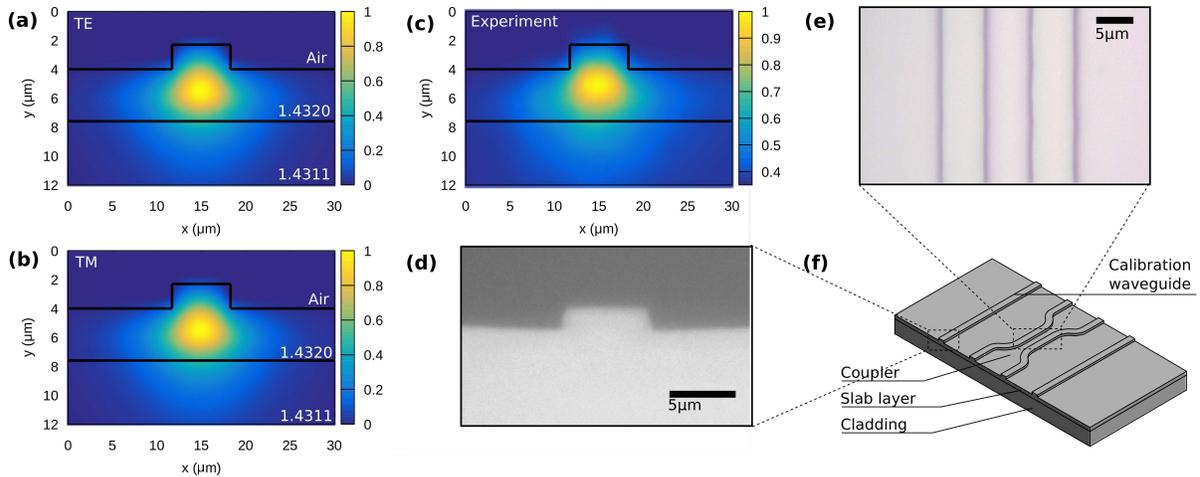}
    \caption{\label{fig:design}(a) and (b) Simulated TE and TM modes of the waveguide structure, obtained using the semi-vectorial finite difference beam propagation method (RSOFT BeamPROP) at \SI{630}{\nano\meter}. The modes are very similar, with $n_{eff} =$ 1.431 (TE \& TM), and appear in good agreement with the experimentally obtained mode in (c). (f) The directional coupler chip design, including coupled and uncoupled waveguides. Optical microscope images in (d) and (e) show the coupled region and end-face of a typical device.}
\end{figure*}

We propose that integrated photonics platforms that can be mechanically tuned may be a solution which will allow rapid tunability while relaxing the existing constraints on wavelength and polarization modes. Such a platform can be constructed on a flexible polymer material such as polydimethyl siloxane (PDMS). Single mode slab waveguides have been fabricated on PDMS chips by using a modified soft imprint lithographic technique (see Figure~\ref{fig:fabrication}), with the necessary refractive index contrast between slab and substrate achieved by modifying the density of the cured polymer\cite{Cai2013,Kee2009}.

Systems of waveguides fabricated in this manner have previously been proposed as solutions for coupling separate optical systems in a robust yet flexible manner\cite{Missinne2014,Kee2010}. In these studies, circuits were designed to be robust under mechanical deformation of the PDMS chip. 
We propose that by taking the opposite approach, it is possible to design devices whose optical properties are responsive to mechanical distortions of the substrate. While previous work has also leveraged an elastomer substrate for tunability~\cite{Chen2012}, our approach relies solely on PDMS microstructures, greatly reducing fabrication complexity. If successful, such a platform would enable a new degree-of-freedom for tuning integrated waveguide platforms.

In this paper, we describe the fabrication and demonstration of directional couplers, which are able to act as beamsplitters with continuously tunable reflectivity. The mechanical tunability is achieved by changing the separation of co-propagating optical modes, with corresponding changes to their coupling coefficient. As previously demonstrated in single mode optical fiber coupler devices~\cite{Digonnet1982}, this can result in a wide range of tunability. By implementing these elements in a single monolithic platform, we pave the way towards more complicated optical circuits with mechanically adjustable behaviours. This platform is compatible with a wide range of visible and near-infrared wavelengths. In addition, it exhibits very similar propagation coefficients for both TE and TM polarization modes, both of which are important considerations for the quantum photonics community.

As a low cost, non-reactive transparent material PDMS has become a workhorse tool in microfluidic, ``lab on a chip'' environments and consequently there is a large body of literature concerning the fabrication of microscale structures. To fabricate our single mode waveguides we adopted a casting technique similar to those used in soft imprint lithography. The liquid prepolymer is introduced to a patterned substrate before curing, and later removed by a peeling process. This approach has been shown to be suitable for the production of micron-scale ribs and trenches with high aspect ratios and low surface roughness.

Our fabrication process is outlined in detail in Figure~\ref{fig:fabrication}. It yields a two layer structure in which a thin layer of PDMS (\SI{4}{\micro\meter}) is patterned with shallow ridges (\SI{1.7}{\micro\meter}$\times$\SI{6.6}{\micro\meter}) to define weakly confined waveguide modes within the slab. A thicker (\SIrange{1}{2}{\milli\meter}) cladding layer provides refractive index contrast and doubles as the mechanical interface for handling and interfacing with the chip. Refractive index contrast of approximately $\Delta n = 0.001$ is achieved by curing the thin slab layer at higher temperature than the cladding layer. The resulting optical modes are shown in Figure~\ref{fig:design}.

We simulate the modes of our waveguides using an iterative solver based on the finite difference beam propagation technique (RSOFT BeamPROP). Generally, quantum photonics experiments prefer that the modes for horizontal and vertically polarized light (usually referred to a TE and TM respectively) be as similar as possible, and indeed this requirement is a key motivator for pursuing platforms other than the commercial silicon photonics systems. Our simulations show propagation coefficients of $n_{eff} = $ 1.431 for both TE and TM, with a relative difference of just 4.2$\times 10^{-6}$. The associated 2D intensity profiles are shown in Figure~\ref{fig:design}(a) and \ref{fig:design}(b) alongside an experimentally obtained mode \ref{fig:design}(c) launched with an elliptical polarization after propagation through a fiber. All modes are in good agreement, with the experimental and simulated mode full width half maxima differing by just 4.7\% and 2.7\% in $x$ and $y$ respectively.

With the waveguides characterized, we proceeded to fabricate beamsplitters using directional couplers such as the one shown in Figure~\ref{fig:design}(e,f). We can use coupled mode theory~\cite{Lifante2003} to predict the behaviour of our device under elastic deformation of the substrate. As the directional coupler is symmetric, the power in the input and output arms (labeled $P_A$ and $P_B$ respectively) will be given after a co-propagation distance $z$ by

\begin{subequations}
\label{eqn:power}
\begin{equation}
P_A(z) = 1-\sin^2(\kappa z)
\end{equation}
\begin{equation}
P_B(z) = \sin^2(\kappa z)
\end{equation}
\end{subequations}

where $\kappa$ is the coupling coefficient, in general related to the convolution of the spatial modes of each waveguide~\cite{Lifante2003}. For example, if the modes are precisely Gaussian we would expect $\kappa$ to also be Gaussian with respect to waveguide separation. We can compute the correlation of our simulated mode with a displaced copy, giving the expected scaling between $\kappa$ and waveguide separation.

\begin{figure}[t]
	\includegraphics[width=\linewidth]{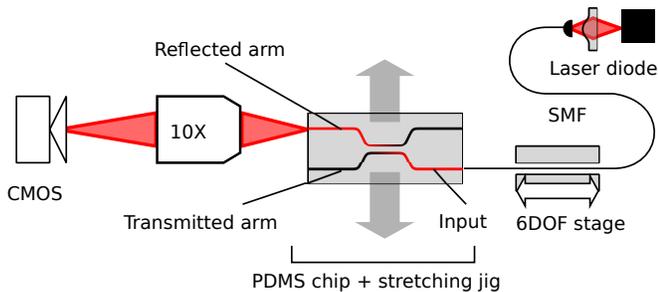}
    \caption{\label{fig:measurement}A schematic of our measurement setup, in which a PDMS chip containing a directional coupler is stretched perpendicular to the propagation direction. Light is coupled into the chip using a single mode optical fiber (SMF) held by a six degree of freedom translation stage (6DOF stage), and the output modes are imaged onto a CMOS sensor via a 10X objective lens.}
\end{figure}

After suitable mounting in a stretching jig driven by a miniature translation stage, we subjected a device containing a directional coupler to a variety of linear deformations in the direction transverse to waveguide propagation, and characterized the response (see Figure~\ref{fig:measurement}). Light was introduced to a directional coupler using butt-coupling from single mode optical fiber, and the output face of the device was imaged onto a CMOS camera (Point Grey Chameleon3) using a 10X objective lens (see Figure~\ref{fig:measurement}). The camera's response to intensity was characterized, after which appropriate treatment of the digital images enabled the relative power in the output modes to be calculated.

\begin{figure}
	\includegraphics[width=0.85\linewidth]{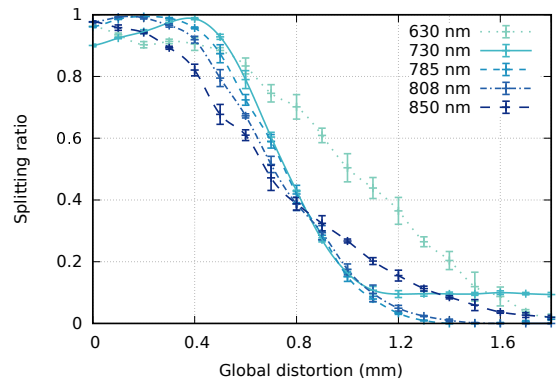}
    \caption{\label{fig:wavelengths}Splitting ratio (defined as $P_A/(P_A+P_B)$) is shown for a range of wavelengths from \SIrange{630}{850}{\nano\meter}. A full range of splitting ratios with high contrast are observed between \SI{785}{\nano\meter} and \SI{850}{\nano\meter}, with the relationship between deformation and splitting ratio differing as expected due to wavelength dependence in the spatial distribution of the optical mode. The device is still serviceable at shorter wavelengths (\SI{730}{\nano\meter} and \SI{630}{\nano\meter}) though with a reduced range of splitting ratios. }
\end{figure}

The directional couplers are robust, with reliable operation over many cycles of stretching and relaxation. We are able to confirm a linear relationship between the global deformation and the change in mode separation at the output face, with no discernible hysteresis. Due to the weakly confining nature of our waveguides, the platform supports single mode operation over a wide wavelength range, and we observe tunable splitting ratios for a range of wavelengths between \SI{630}{\nano\meter} and \SI{850}{\nano\meter}, with qualitatively similar results (see Figure~\ref{fig:wavelengths}).

\begin{figure}
	\includegraphics[width=0.85\linewidth]{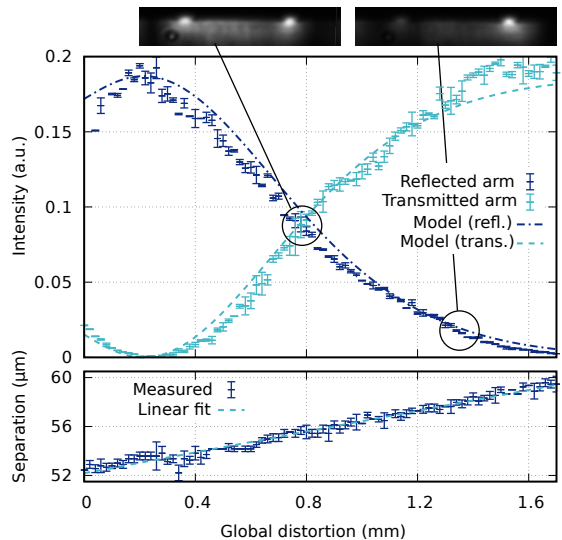}
    \caption{\label{fig:850stretch}The normalized power in the reflected and transmitted arms (see Figure~\ref{fig:measurement}) as a function of chip-scale distortion, measured at \SI{850}{\nano\meter}. The centre-to-centre distance of the modes was also calculated, indicating an approximately linear deformation of the directional coupler under test by around \SI{5}{\micro\meter}. Over this range of deformations the device is tuned from approximately 99.9\% reflecting to approximately 98.5\% transmitting. The relationship between the deformation and the power in the transmitted and reflected modes is fit using Equation~\ref{eqn:power}, with the $\kappa$ parameter calculated from numerical simulation of the mode at \SI{850}{\nano\meter}. Two representative microscope images of the output face are also shown (inset), corresponding to approximately equal mode power and to a highly distorted state, with most power concentrated in the transmitted arm.}
\end{figure}

Figure~\ref{fig:850stretch} shows the behaviour of a typical directional coupler as the chip is stretched. We observe initially very high optical coupling to the second waveguide, which we term ``reflectance'' by analogy to bulk beamsplitters. The ratio of reflected light to transmitted light falls off as the deformation is increased. The trend in output mode power appears well approximated by the retarded cosine described in equations~\ref{eqn:power}, where the relationship between deformation and the parameter $\kappa$ is computed from numerically obtained modes. Data collected at \SI{850}{\nano\meter} indicates an extinction ratio of 99.9\% for the transmitted arm and 98.8\% for the reflected arm.


To conclude, we have demonstrated a tunable beamsplitter based on directional couplers in a flexible, polymer-based waveguide platform. The beamsplitter shows wide tunability over a range of wavelengths, enabling nearly the full range of splitting ratios from \SI{630}{\nano\meter} to \SI{850}{\nano\meter}. The platform is low cost and suitable for rapid prototyping, and uses fabrication techniques which are amenable to future  scalability~\cite{Hassanin2011}. 

While we have demonstrated sensitivity to large scale deformation of the chip substrate, the operating principle should also permit tunability via local deformations, for example by applying point-like pressure from below and deforming the structure out-of-plane. We are currently working to extend our devices to enable addressing in this way, resulting in independent control of multiple directional couplers. This basic tunable optical component may then be adopted as a building block in larger devices with many input and output modes.

With low sensitivity to the input polarization, we believe large arrays of these devices will find application in quantum photonic simulation experiments, enabling a range of circuits to be investigated using a single device compatible with polarization entangled photons at visible and near-infrared wavelengths. For example, it is possible to fabricate large scale linear arrays, with the resulting photonic random walks tuned by adjusting the mode coupling via mechanical deformations of the host chip. We also envisage devices capable of performing polarization mode or wavelength multiplexing for (quantum) communication schemes \cite{vest15}.


Looking further ahead, we believe the inclusion of optical nonlinearity in this platform will greatly enhance its usefulness. This could be achieved by infusion of micro- or nano-particles into the liquid PDMS precursor, by chemical modification of the cured polymer or by decorating the PDMS microstructures with appropriate materials \cite{koperskia17}. In this way it may become possible to tune both the linear and nonlinear optical properties of the system by mechanical means.

\section*{Acknowledgements}

This work is supported by Singapore Ministry of Education Academic Research Fund Tier 3 (Grant No. MOE2012-T3-1-009).

\bibliography{pdms}

\end{document}